\documentclass[prd,onecolumn,showpacs,nofootinbib,preprintnumbers,superscriptaddress]{revtex4}
\usepackage{amsmath}
\usepackage{amsfonts}
\usepackage{graphicx}
\usepackage{hyperref}
\usepackage{mathbbol}

\def\be{\begin{equation}}
\def\ee{\end{equation}}
\def\bea{\begin{eqnarray}}
\def\eea{\end{eqnarray}}
\def\f{\frac}
\def\s{\sqrt}
\def\l{\left}
\def\r{\right}
\def\e{\epsilon}

\def\k{\kappa}
\def\no{\nonumber}
\begin{document}

\title{Phantom
expansion with non-linear Schr\"{o}dinger-type formulation of scalar
field cosmology}

\author{Theerakarn Phetnora}\email{u48313065@nu.ac.th}
\address{Fundamental Physics \&
Cosmology Research Unit\\ The Tah Poe Academia Institute (TPTP),
Department of Physics\\ Naresuan University, Phitsanulok 65000,
Siam}
\author{Roongtum Sooksan}
\address{Fundamental Physics \&
Cosmology Research Unit\\ The Tah Poe Academia Institute (TPTP),
Department of Physics\\ Naresuan University, Phitsanulok 65000,
Siam}

\author{Burin Gumjudpai \footnote{Corresponding author}}\email{buring@nu.ac.th} \email{B.Gumjudpai@damtp.cam.ac.uk}
\address{Fundamental Physics \&
Cosmology Research Unit\\ The Tah Poe Academia Institute (TPTP),
Department of Physics\\ Naresuan University, Phitsanulok 65000,
Siam}
\address{Centre for Theoretical Cosmology, Department of Applied
Mathematics and Theoretical Physics\\ University of Cambridge,
Centre for Mathematical Sciences\\ Wilberforce Road,
 Cambridge CB3 0WA, United Kingdom}

\begin{abstract}
 We describe non-flat standard Friedmann cosmology of canonical scalar field with barotropic fluid in form of non-linear
  Schr\"{o}dinger-type (NLS) formulation in which all cosmological dynamical quantities are expressed in term of
Schr\"{o}dinger quantities as similar to those in
 time-independent quantum mechanics. We assume the expansion to be superfast, i.e. phantom expansion. We
report all Schr\"{o}dinger-analogous
 quantities to scalar field cosmology.
  Effective equation of state coefficient is analyzed and
 illustrated. We show that in a non-flat universe, there is no fixed $w_{\rm eff}$ value for the phantom divide. In a non-flat universe, even $w_{\rm eff}
> -1$, the expansion can be phantom. Moreover, in open universe, phantom
expansion can happen even with $w_{\rm eff}
> 0$. We also report scalar field exact solutions within frameworks of the Friedmann
formulation and the NLS formulation in non-flat universe cases.
\end{abstract}


\date{\today}

\vskip 1pc

\maketitle \vskip 1pc

\section{Introduction}

\label{sec:introduction}
 Supernovae Type Ia data and cosmic microwave background observations show recently strong evidence of present accelerating phase of the universe
 \cite{Masi:2002hp,Scranton:2003in,Riess:1998cb,Riess:2004nr,Astier:2005qq} while nowadays inflationary paradigm in the early universe is one
 of the corner stones in cosmology \cite{inflation}.
Present acceleration and inflation of the universe are both believed
to result from effect of either dynamical scalar field with
time-dependent equation of state coefficient $w_{\phi}(t) < -1/3$ or
a cosmological constant with $w = -1$. Alternative explanation for
present acceleration to dark energy is modification of general
relativity which includes braneworld models (for review, see
\cite{Nojiri:2006ri} and references therein). Among these ideas, the
scalar field catches most attention therefore many analysis in
cosmological contexts and observations have been carried out
\cite{Padmanabhan:2004av}. Conventional formulation of canonical
scalar field cosmology with barotropic perfect fluid, can also be
expressed as non-linear Ermakov-Pinney equation as shown recently
\cite{Hawkins:2001zx,Williams:2005bp}. However, non-Ermakov-Pinney
equation for such system can also be written in form of a non-linear
Schr\"{o}dinger-type equation (NLS). The solutions of the NLS-type
equation correspond to solutions of the generalized Ermakov-Pinney
equation of scalar field cosmology
\cite{Williams:2005bp,D'Ambroise:2006kg}. The NLS-type formulation
was concluded and shown in case of power-law expansion in Ref.
\cite{Gumjudpai:2007qq} where all Schr\"{o}dinger-type quantities
corresponding to scalar field cosmology are worked out. NLS-type
formulation also provides an alternative way of solving for the
scalar field exact solutions in various cases even with non-zero
curvature \cite{Gumjudpai:2007bx}.

Various observations allow scalar field equation of state coefficient, $w_{\phi}$ to be less than -1 \cite{Melchiorri:2002ux}. Previous evidence
from combined cosmic microwave background, large scale structure survey and supernovae type Ia without assuming flat universe yields $ w_{\phi}
= -1.06^{+0.13}_{-0.08}\;$\cite{Spergel:2006hy} while using supernovae data alone assuming flat universe yields $w_{\phi}=-1.07\pm
0.09\;$\cite{WoodVasey:2007jb}. The most recent WMAP five-year result \cite{Hinshaw:2008kr,Dunkley:2008ie} combined with Baryon Acoustic
Oscillation (BAO) of large scale structure survey: SDSS and 2dFGRS\cite{Percival:2007yw} and type Ia supernovae data from
HST\cite{Riess:2004nr}, SNLS\cite{Astier:2005qq} and ESSENCE\cite{WoodVasey:2007jb} assuming dynamical $w$ with flat universe yields $-1.33 <
w_{\phi,0} < -0.79$ at 95\% confident level \cite{Komatsu:2008hk}. Also this data with additional BBN constraint of limit of expansion rate
\cite{Steigman:2007xt,Wright:2007vr} yields $-1.29 < w_{\phi,0} < -0.79$ at 95\% confident level and $w_{\phi,0} = -1.04 \pm 0.13$ at 68\%
confident level \cite{Komatsu:2008hk}. This suggests that the scalar field could be phantom, i.e $w_{\phi}<-1$ \cite{Caldwell:1999ew}. For a
canonical scalar field, phantom behavior can be attained by negative kinetic energy term of the scalar field Lagrangian density. In FLRW general
relativistic cosmology, there is a Big Rip singularity with $a, \rho, |p| \rightarrow \infty$ at finite time \cite{Caldwell:2003vq},
nevertheless singularity avoidance has been attempted in various ways \cite{Sami:2005zc}. Extension to include phantom field case in NLS-type
formulation was made in \cite{Gumjudpai:2007qq}. In NLS-type formulation one can presume any law of expansion $a=a(t)$, e.g. power law $a \sim
t^q$ or exponential expansion, $a\sim \exp(t/\tau)\;$ \cite{Gumjudpai:2007qq,Gumjudpai:2007bx} and works out all NLS-quantities keeping open
possibility for the field to be phantom or non-phantom and non-zero spatial curvature. Analogous studies to the slow-roll, WKB and the Big Rip
in NLS formulation were done in \cite{Gumjudpai:2008}.

 To attain accelerating
expansion, one needs to have effective equation of state
coefficient, $w_{\rm eff} < -1/3$ where \be  w_{\rm eff} =
\f{{\rho_{\phi}w_{\phi} + \rho_{\gamma} w_{\gamma}}}{{\rho_{\rm
tot}}}\,, \label{weff}\ee $\rho_{\gamma}$ is density of barotropic
fluid, $\rho_{\phi}$ is density of the scalar field and $\rho_{\rm
tot} = \rho_{\phi}+\rho_{\gamma}$. It has been known in standard
cosmology that for flat universe ($k=0$), if the expansion is $a
\sim t^q $ , then $-1 < w_{\rm eff} < -1/3$; if $a \sim \exp(t/\tau)
$ , then $ w_{\rm eff}= -1$ and if $a \sim (t_{\rm a} -t)^q $ , then
$ w_{\rm eff} < -1$. Here $q \equiv 2/[3(1+w_{\rm eff})]$, $\tau,
t_{\rm a}$ are finite characteristic times. In the last case, $
w_{\rm eff} < -1$ corresponds to $q<0$.

In this work, we consider phantom expansion $a \sim (t_{\rm a} -t)^q
$ in the NLS-type formulation with non-zero curvature $k$. We
introduce cosmological system in Sec. \ref{sec:System}, then
NLS-type formulation in Sec. \ref{sec:NS}. The Schr\"{o}dinger
quantities for phantom
 expansion are presented in Sec. \ref{sec:phantom} where we
analyze value of $w_{\rm eff} $ and show
 conditions of how much negative $w_{\phi}$ must be in order to keep
 the expansion phantom. We also illustrate parametric plots for $w_{\rm
 eff}$ with $q$ and $t$. Scalar field exact solutions solved from
 both standard formulation and NLS-type formulation are given in Sec.
 \ref{sec:solution} where we comment on both procedures of
 obtaining the solutions. Finally we
 conclude our work in Sec. \ref{sec:conclusion}.

\section{Cosmological System} \label{sec:System}

Barotropic fluid and scalar field fluid are major components in our
scenario. The perfect barotropic fluid pressure $p_{\gamma}$ and
density $\rho_{\gamma}$ obey an equation of state, $p_{\gamma} =
(\gamma-1)\rho_{\gamma} = w_{\gamma}\rho_{\gamma}$ while for scalar
field, $p_{\phi} = w_{\phi}\rho_{\phi}$. Total density and total
pressure are $\rho_{\rm tot} = \rho_{\gamma} + \rho_{\phi}$ and
$p_{\rm tot} = p_{\gamma} + p_{\phi}$. For the barotropic fluid,
$w_{\gamma}$ is written in term of $n$. We set $w_{\gamma} \equiv
(n-3)/3$ so that $n = 3(1+w_{\gamma}) = 3\gamma$,  hence
$w_{\gamma}=-1$ corresponds to $n=0$, $w_{\gamma}=-1/3$ to $n=2$,
$w_{\gamma}=0$ to $n=3$, $w_{\gamma}=1/3$ to $n=4$, and
$w_{\gamma}=1$ to $n=6$. The conservation equation is hence \bea
\dot{\rho}_{\gamma} = - n H \rho_{\gamma}\,, \label{1} \eea
 with
solution,
\be \rho_{\gamma} = \frac{D}{a^{n}}\,.\label{barorho} \ee %
Therefore $ p_{\gamma} = [{(n-3)}/{3}]({D}/{a^{n}})\,, $ where $a$
is scale factor, the dot denotes time derivative, $H = \dot{a}/a$ is
Hubble parameter and $D\geq 0$ is a proportional constant. We
consider scalar field that is minimally coupling to gravity with
Lagrangian density, $ \mathcal{L} = (1/2)\epsilon \dot{\phi}^2 -
V(\phi)\,, $ of which $\epsilon=1$ for non-phantom case and $-1$ for
phantom case. Density and pressure of the field are given as
\bea \rho_{\phi} = \frac{1}{2} \epsilon \dot{\phi}^2 + V(\phi)\,,
\;\;\;\;\;\;\;\;\;\; p_{\phi}= \frac{1}{2} \epsilon \dot{\phi}^2 -
V(\phi)\,, \label{phanp}\eea therefore \bea w_{\phi} =
\f{p_{\phi}}{\rho_{\phi}} = \f{\epsilon \dot{\phi}^2
-2V(\phi)}{\epsilon \dot{\phi}^2 + 2V(\phi)}\,. \label{wphi}\eea The
field obeys conservation equation
\be \epsilon\left[\ddot{\phi} + 3H\dot{\phi} \right] + \frac{{\rm
d}V}{{\rm d}\phi} = 0\,. \label{phanflu} \ee
Considering
Friedmann-Lema\^{i}tre-Robertson-Walker (FLRW) universe, the
Friedmann equation and acceleration equation are
\bea H^2 &=& \frac{\kappa^2}{3}\rho_{\rm tot} - \frac{k}{a^2}\,,
\label{fr}
\\ \frac{\ddot{a}}{a} &=&  -\frac{\kappa^2}{6} \rho_{\rm tot}
(1 + 3w_{\rm eff}) \,,  \label{ac}\eea %%
where $\kappa^2 \equiv 8\pi G = 1/M_{\rm P}^2$, $G$ is Newton's
gravitational constant, $M_{\rm P}$ is reduced Planck mass and $k$
is spatial curvature. Using Eqs. (\ref{barorho}), (\ref{phanp}),
(\ref{phanflu}) and (\ref{fr}), one can show that
\bea   \epsilon \dot{\phi}(t)^2 & = & -\frac{2}{ \kappa^2} \left[ \dot{H} - \frac{k}{a^2}  \right] - \frac{n D}{3  a^n} \,, \label{phigr} \\
V(\phi) &=& \frac{3}{\kappa^2} \left[H^2 + \frac{\dot{H}}{3} +
\frac{2k}{3 a^2} \right] + \left(\frac{n-6}{6}\right)
\frac{D}{a^n}\,. \label{Vgr} \eea %

\section{NLS-type formulation}  \label{sec:NS}
Non-linear Schr\"{o}dinger-type formulation for canonical scalar
field cosmology and  barotropic fluid was proposed by
 J.~D'Ambroise and F.~L.~Williams\cite{D'Ambroise:2006kg} and was also extended to include phantom field
case\cite{Gumjudpai:2007qq}\footnote{It is worth noting that Schr\"{o}dinger-type equation in scalar field cosmology was previously considered
in different procedure to study inflation and phantom field problems \cite{Chervon:1999}.}. In the Schr\"{o}dinger formulation, wave function
$u(x)$ is related to scale factor in cosmology as
 \bea
 u(x) &\equiv& a(t)^{-n/2}\,, \label{utoa} \eea
 while Schr\"{o}dinger total energy $E$ and Schr\"{o}dinger potential
 $P(x)$ are linked to cosmology as
 \bea
 E &\equiv&  -\frac{\kappa^2 n^2}{12} D \,, \label{E} \\
P(x) &\equiv& \frac{\kappa^2 n}{4}a(t)^{n} \epsilon \dot{\phi}(t)^2
\,. \label{schropotential} \eea
These quantities satisfy a non-linear Schr\"{o}dinger-type equation:
 \bea \frac{{\rm d}^2 }{{\rm d}x^2}u(x) + \left[E-P(x)\right]
u(x)
 = -\frac{nk}{2}u(x)^{(4-n)/n}\,,   \label{schroeq} \eea
with a mapping from $t$ to $x$ is via \be x = \sigma(t) \label{xt},
\ee such that \cite{Gumjudpai:2007qq,Gumjudpai:2007bx}
\bea \dot{x}(t)&=& u(x)\,, \label{dsigtou} \\
\phi(t) &=& \psi(x)\,= \frac{\pm
2}{\kappa\sqrt{n}}\int{\sqrt{\frac{P(x)}{\epsilon}}}\,{\rm d}x\,.
\label{psi} \eea
If $P(x) \neq 0$ and $n \neq 0$, inverse function of $\psi(x)$
exists as $\psi^{-1}(x)$. Therefore $ x(t) = \psi^{-1}\circ \phi(t)
$ and the scalar field potential, $V\circ \sigma^{-1}(x)$
can be expressed as, %
\be V(t) = \frac{12}{\kappa^2 n^2}\left( \frac{{\rm d} u}{{\rm d}x}
\right)^2 - \frac{2 u^2}{\kappa^2 n} P(x) + \frac{12 u^2}{\kappa^2
n^2}E + \frac{3 k u^{4/n}}{\kappa^2} \,. \label{vt} \ee

\section{Phantom expansion} \label{sec:phantom}
Expansion of the form $a \sim (t_{\rm a} -t)^q $ is called phantom
when $q < 0$ for a flat universe. Here in non-flat universe, $q$ is
considered to possess any value and the term phantom expansion also
refers to expansion function of the form $a \sim (t_{\rm a} -t)^q $
as in the flat case.
\subsection{NLS-type formulation for phantom expansion}
\begin{figure}[t]
\begin{center}
\includegraphics[width=6.5cm,height=6.4cm,angle=0]{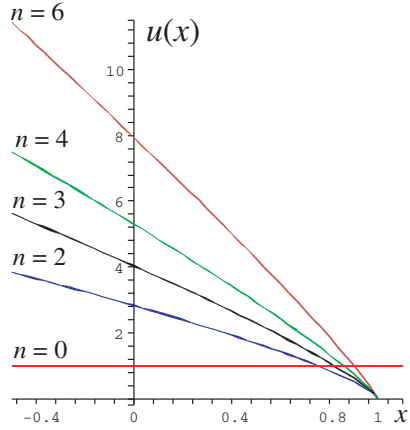}
\end{center}
\caption{Schr\"{o}dinger wave function, $u(x)$ when assuming phantom
expansion. $u(x)$ depends on only $q$, $n$  and $t_{\rm a}$ but does
not depend on $k$. Here we set $t_{\rm a} =1.0$ and $q = -6.666$. If
$k=0$, $q=-6.666$ corresponds to $w_{\rm eff} = -1.1$.
\label{figux}}
\end{figure}

\begin{figure}[t]
\begin{center}
\includegraphics[width=8.1cm,height=11.9cm,angle=0]{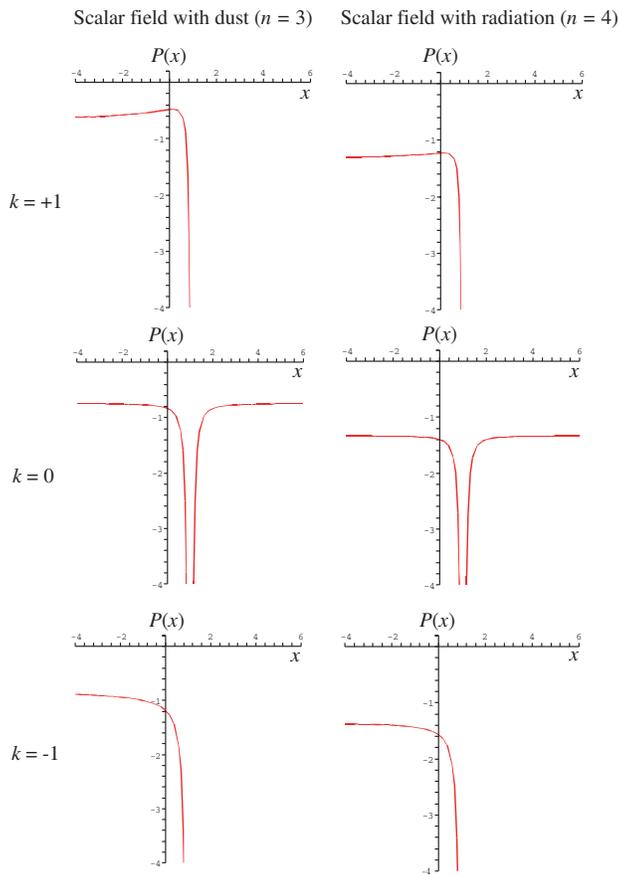}
\end{center}
\caption{Schr\"{o}dinger potential in phantom expansion case for
dust and radiation fluids with $k= 0, \pm1$. Numerical parameters
are as in the $u(x)$ plots (Fig. \ref{figux}). $x_0$ is set to $1$.
For non-zero $k$, there is only one real branch of $P(x)$.
\label{figpx}}
\end{figure}

With the phantom expansion, $a \sim (t_{\rm a} -t)^q $, we use
 Eqs. (\ref{utoa}) and
(\ref{dsigtou}) to relate
Schr\"{o}dinger wave function to standard cosmological quantity as %
\bea
u(x) & = & \dot{x}(t) = (t_{\rm a} - t)^{-qn/2} \,. \label{uast}
\eea Integrate the equation above so that the Schr\"{o}dinger scale,
$x$
is related to cosmic time scale, $t$ as %
\bea x(t) & = & \frac{1}{\beta}\,(t_{\rm a} - t)^{-\beta} + x_0\,,
\label{xast}
\eea where $\beta \equiv (qn-2)/2 $ and $x_0$ is an integrating constant. Conversely, %
 \bea
t(x) & = &  t_{\rm a} - \frac{1}{\left[ \beta (x-x_0)  \right]^{1/\beta}} \,. \label{tasx} \eea \\
The  Schr\"{o}dinger wave function can be directly found from Eqs.
(\ref{uast}) and (\ref{tasx}) as %
\bea u(x) & = & \left[ \beta (x-x_0) \right]^{qn/(qn-2)}\,.%
 \eea
For $a \sim (t_{\rm a} -t)^q $,
we can find %
$\epsilon\dot{\phi}(t)^2$ from Eq. (\ref{phigr}): %
\bea \epsilon\dot{\phi}(t)^2 & = & \frac{2q}{\kappa^2 (t_{\rm a} -
t)^2} + \frac{2k}{\kappa^2 (t_{\rm a} -t)^{2q}}  - \frac{nD}{3
(t_{\rm a} - t)^{qn}}\,. \label{phit}  \eea Using (\ref{phit}) with
phantom expansion in Eq. (\ref{schropotential}), therefore
 \bea
 P(t) & = & \frac{qn}{2} (t_{\rm a} -t)^{qn-2}
+\frac{kn}{2}(t_{\rm a} - t)^{q(n-2)} - \frac{\kappa^2 n^2
D}{12}\,,  \eea %
which can be expressed in term of $x$ using Eq. (\ref{tasx}) as \bea
 P(x) = \frac{2qn}{(qn-2)^2}\frac{1}{(x-x_0)^2}
 + \frac{kn}{2}\left[ \frac{2}{(qn-2)(x-x_0)}
\right]^{2q(n-2)/(qn-2)}  -  \frac{\kappa^2 n^2 D }{12}. \no \\
\label{Pxphantom}
\eea One might have a thought that all functions in phantom
expansion case can be changed to those in power-law expansion case
by interchanging $(t_{\rm a}-t) \Leftrightarrow t$. However when
$(t_{\rm a}-t)$ is differentiated, there is an extra minus sign. The
Eq. (\ref{Pxphantom}) slightly defers from that of the power-law
expansion case because in the power-law case, the numerator of the
second term is $-2$ instead of $2$. The Schr\"{o}dinger kinetic
energy $T$ is negative value of the first two terms of the
Schr\"{o}dinger potential. At last, the scalar field potential
obtained from Eq.
(\ref{vt}) is %
\be V(t) = \frac{q(3q-1)}{\kappa^2 (t_{\rm a} -t)^2} +
\frac{2k}{\kappa^2 (t_{\rm a} -t)^{2q}} +
\left(\frac{n-6}{6}\right)\frac{D}{(t_{\rm a} -t)^{qn}}\,.
\label{vtphantom}\ee %
which can be checked by using $a \sim (t_{\rm a} -t)^q $ in
(\ref{Vgr}). Wave function of the NLS-formulation is found to be
non-normalizable \cite{Gumjudpai:2007qq} as seen Fig. in \ref{figux}
for case of phantom expansion with various types of barotropic
fluid. Here $q$ is chosen to $-6.666$. In flat universe $q=-6.666$
can be attained when $w_{\rm eff} = -1.1$. Fig. \ref{figpx} shows
$P(x)$ plots for three cases of $k$ with dust and radiation. In
there $x_0 = 1$, therefore $P(x)$ goes to negative infinity at $x =
1$.

\begin{figure}[t]
\begin{center}
\includegraphics[width=6.48cm,height=4.8cm,angle=0]{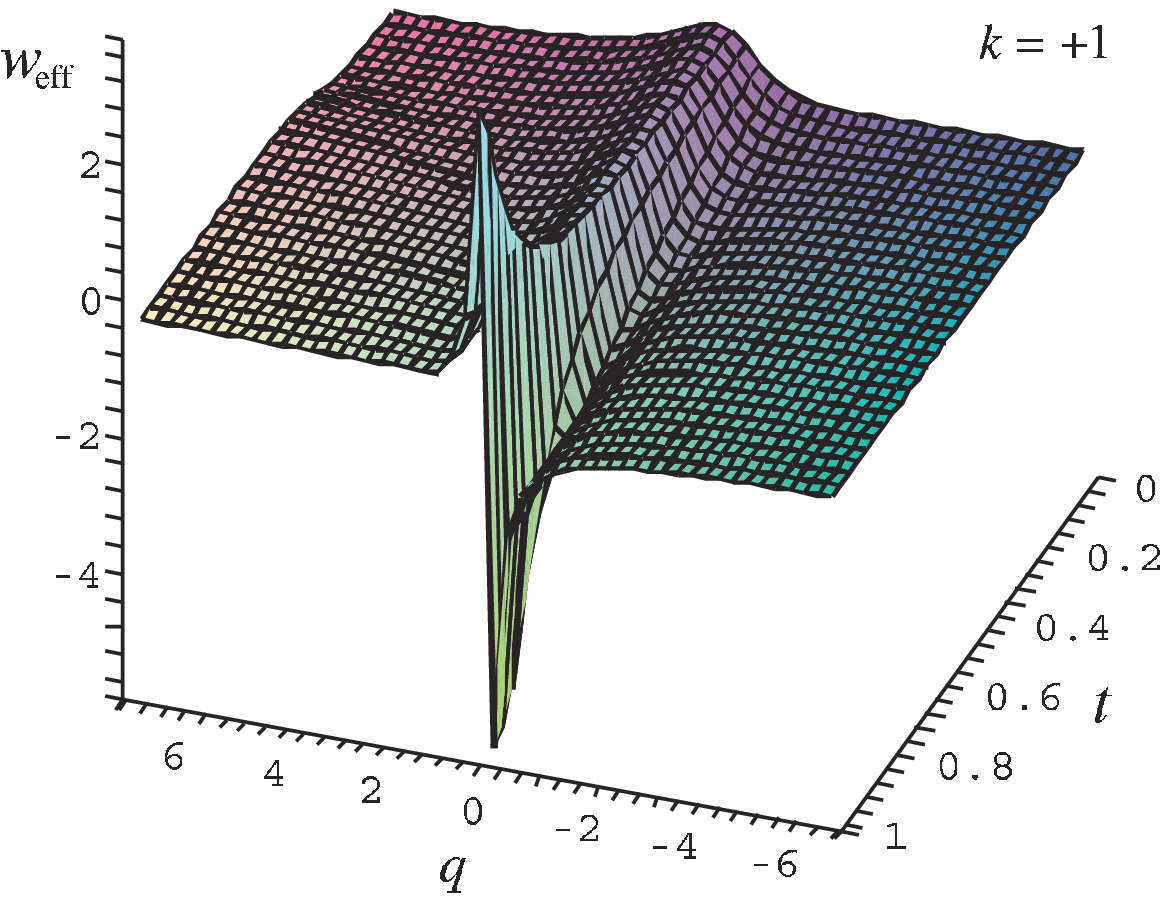}\\
\includegraphics[width=6.48cm,height=5.65cm,angle=0]{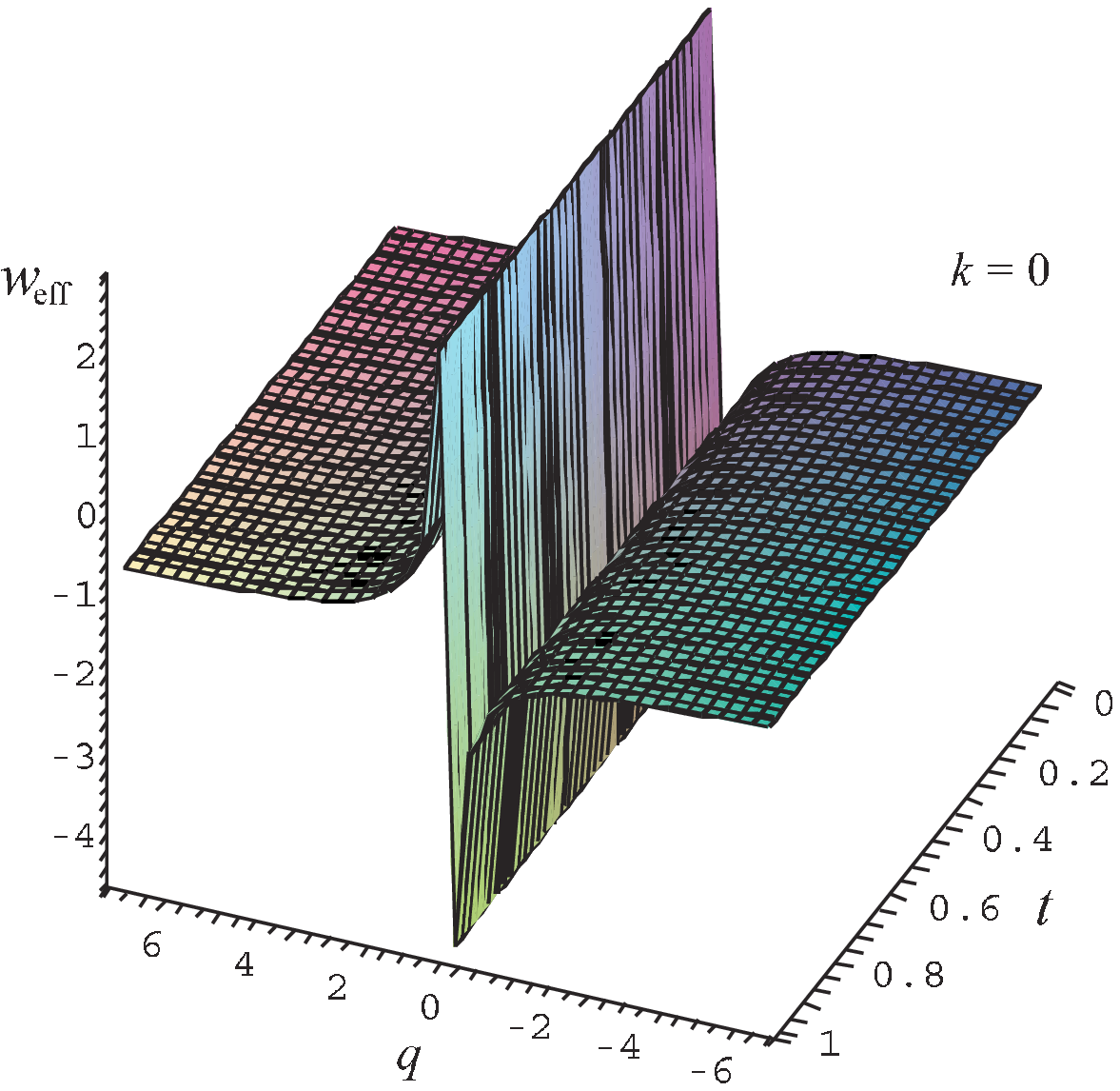}\\
\includegraphics[width=6.48cm,height=5.3cm,angle=0]{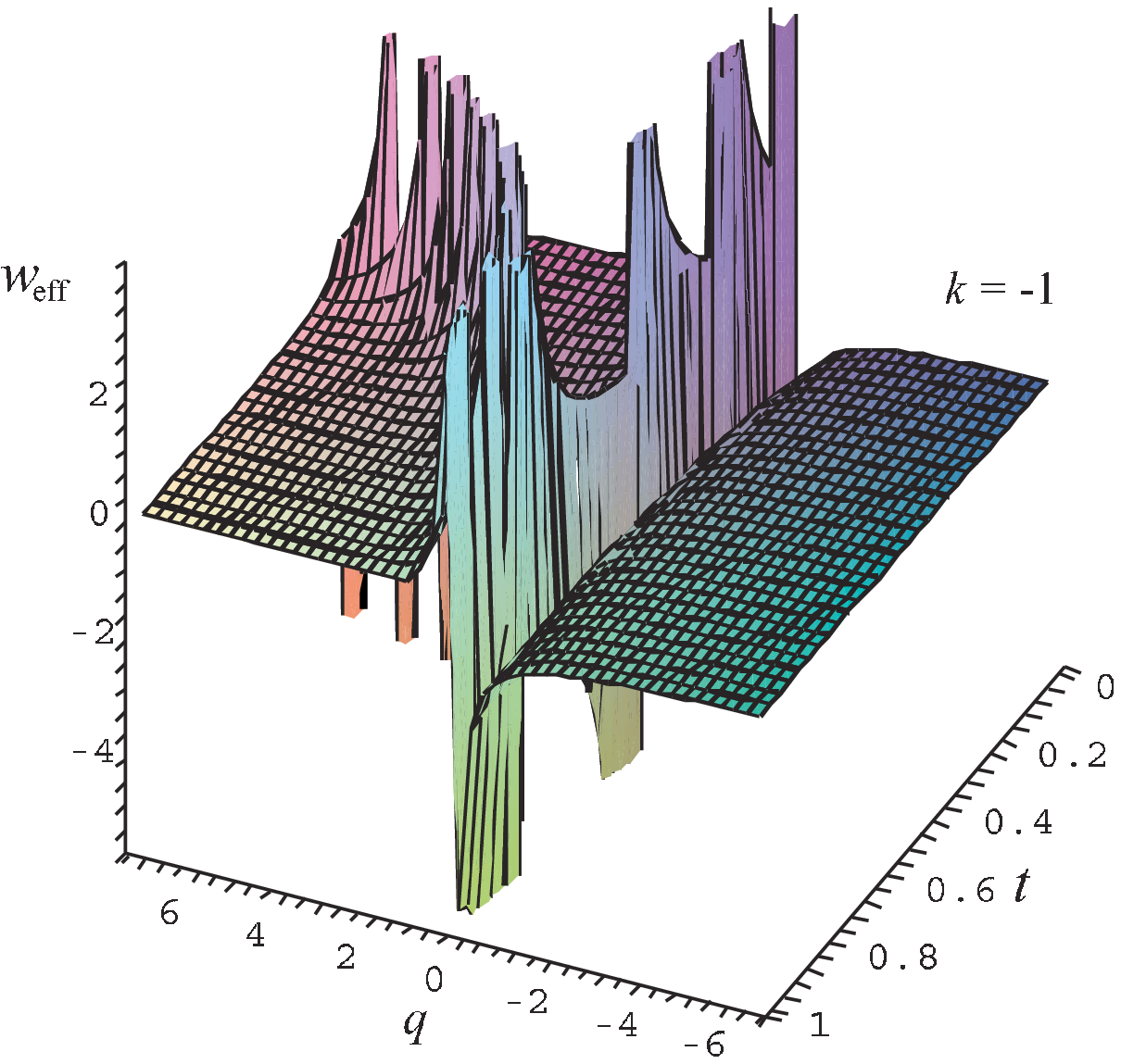}
\end{center}
\caption{Parametric plots of $w_{\rm eff}$ for the expansion $a \sim
(t_{\rm a}-t)^q$ in closed, flat and open universe. Here $t_{\rm a}$
is set to 1. \label{figweff}}
\end{figure}
\begin{figure}[t]
\begin{center}
\includegraphics[width=6.48cm,height=5.83cm,angle=0]{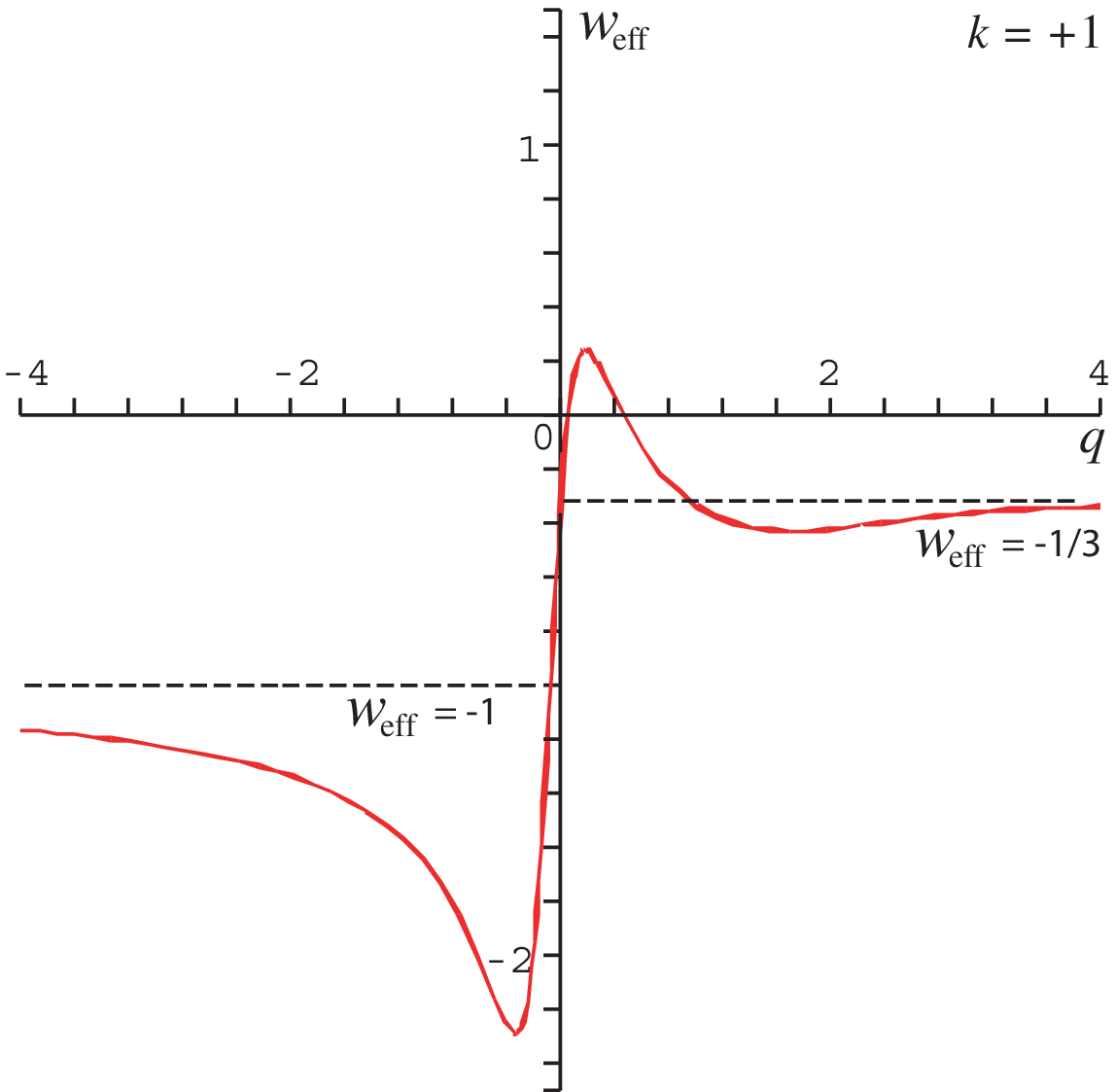}\\
\includegraphics[width=6.48cm,height=5.73cm,angle=0]{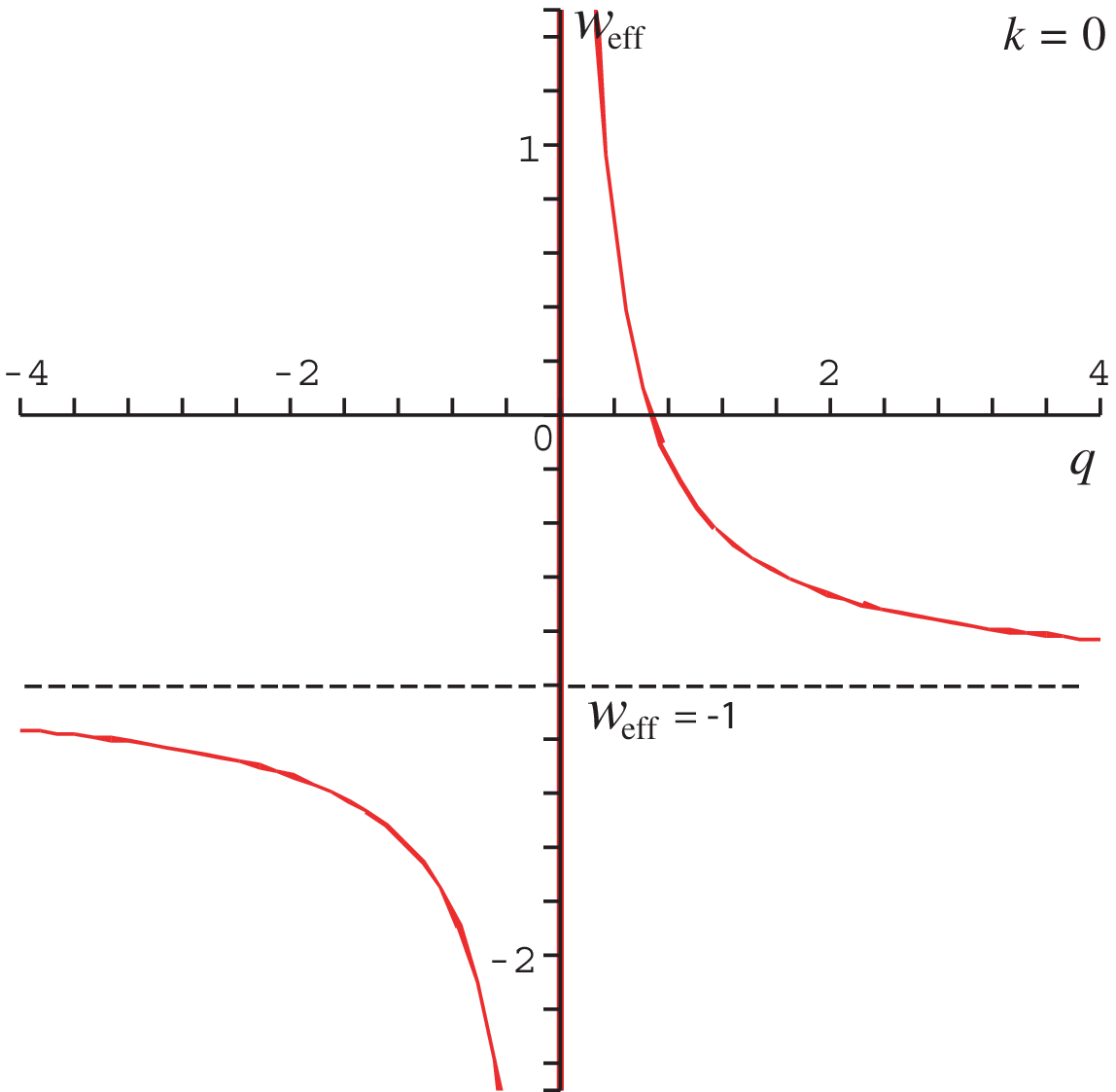}\\
\includegraphics[width=6.48cm,height=5.37cm,angle=0]{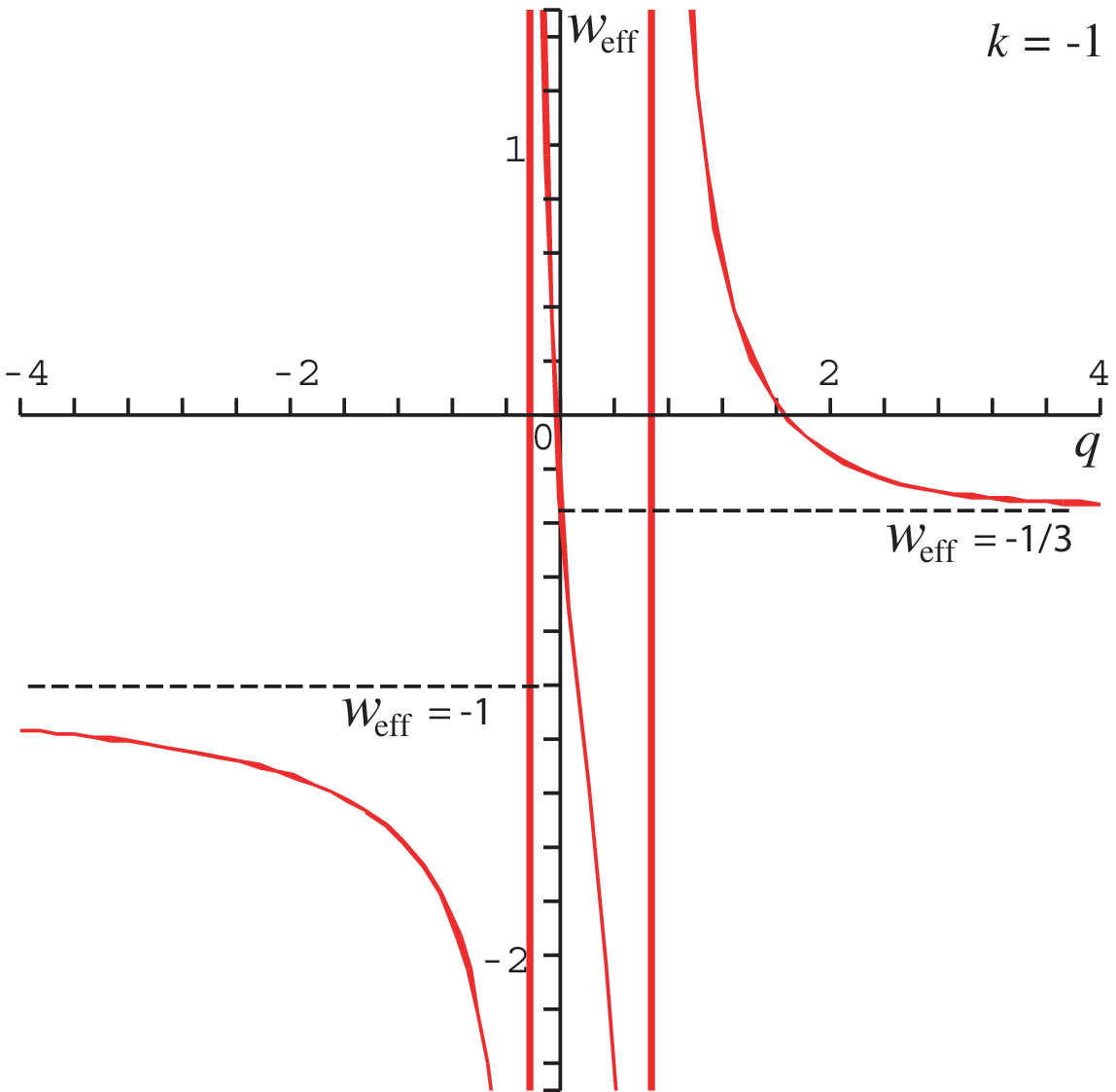}
\end{center}
\caption{$w_{\rm eff}$ for the expansion $a \sim (t_{\rm a}-t)^q$ in
closed, flat and open universe. Here $t_{\rm a}$ is set to 1 and $t$
is 0.7.  \label{fig2Dweff}}
\end{figure}
%%%
\subsection{Analysis on effective equation of state coefficient}
\label{sec:eff}

The definition of effective equation of state coefficient, $ w_{\rm eff} = ({\rho_{\phi}w_{\phi} + \rho_{\gamma} w_{\gamma}})/{\rho_{\rm
tot}}\,$ together with Eq. (\ref{phanp}) and the results in Eqs. (\ref{phit})  and (\ref{vtphantom}) in context of
phantom expansion $a \sim (t_{\rm a} - t)^q $, we can derive %
\be w_{\rm eff} = \frac{(-3q^2 +2q)(t_{\rm a} -t)^{2q-2} - k}{3q^2
(t_{\rm a} -t)^{2q-2} + 3k}\,. \label{Phantomweffkt} %
 \ee
There is a locus, %%
\be
  t = t_{\rm a} - \left(\frac{-k}{q^2}\right)^{1/(2q-2)}\,,
\label{locus}\ee %
where $w_{\rm eff}$ becomes infinite along the locus. Hence for
$k=-1\,$ the locus is $ t = t_{\rm a} - {q^{-1/(q-1)}}$ (in term of
$x$, it is $
  x = [{2}/({qn-2})] q^{(qn-2)/2(q-1)} + x_0 $).
Hence for $k=0$, the coefficient $w_{\rm eff}$ is infinite at $q=0$
or $t=t_{\rm a}$.
  It seems from
the equation above that $w_{\rm eff}$ does not depend on properties,
$n$ or amount of the barotropic fluid, $D$. Indeed $w_{\rm eff}$
implicitly  depends on $D$ and $n$ since time variable and $q$ are
related to $\rho_{\gamma}$ in the Friedmann equation. If $k=0$, it
reduces to $q = 2/3(1+w_{\rm eff})$ and therefore the phantom
condition $w_{\rm eff} < -1$
 implies $q<0$ as it is known.
   This corresponds to a
condition, %
\be w_{\phi} < - 1
-(1+w_{\gamma})\frac{\rho_{\gamma}}{\rho_{\phi}}\,. \ee Therefore
for a fluid with $w_{\gamma} > -1$, $w_{\phi}$ is always less than
$-1$ in a flat universe. In order to have the expansion $a \sim
(t_{\rm a} -t)^q$ in $k=0$ universe, we must have $w_{\rm eff} <
-1$, i.e. in phantom region.
We can rewrite $w_{\phi}$ in term of $w_{\rm eff}$ as %
\be %
w_{\phi} = \frac{    [   \frac{3q^2}{\kappa^2} (t_{\rm a}-t)^{-2} +
\frac{3k}{\kappa^2} (t_{\rm a}-t)^{-2q} ] w_{\rm eff}   -
\frac{n-3}{3} D (t_{\rm a}-t)^{-qn}   }{\frac{3q^2}{\kappa^2}
(t_{\rm a}-t)^{-2}  + \frac{3k}{\kappa^2} (t_{\rm a}-t)^{-2q} -  D
(t_{\rm a}-t)^{-qn}   }\,.  \label{Phantomwphikt}
\ee %
 Eq. (\ref{Phantomwphikt}), when $D=0$ and $k=0$, yields
$w_{\phi} = w_{\rm eff}$. Albeit we set only $D=0$, it gives the same result since $w_{\phi}$ is independent of geometrical background. However,
since the expansion law is fixed, $w_{\phi}$ is tied up with $D$ implicitly via Eq. (\ref{weff}). Note that $w_{\phi}$ has value in the range
$(-\infty,-1]$ and $[1,\infty)$ so that the phantom crossing can not happen when the scalar field is dominant. However,
presence of the dust barotropic fluid in the system gives a
multiplication factor that is less than 1 to the equation of state, i.e.
\be
w_{\rm eff} = \l(\frac{\rho_{\phi}}{\rho_{\phi}+\rho_{\gamma}}\r) w_{\phi}\,.
\ee
We can see that the phantom crossing from $w_{\rm eff} > -1 $ to $w_{\rm eff} < -1$ can happen in this situation.
Fig. \ref{figweff} presents parametric plots of the
 $w_{\rm eff}, q, t$ diagram for various $k$ values.
From the figure, we see the locus in Eq. (\ref{locus}) where $w_{\rm eff}$ blows up. In the parametric plots, the value of $w_{\rm eff}$ at any instance can be obtained
if we know the value of $q$. We need to know $q$ from observation in order to know the realistic value of $w_{\rm eff}$ or the other way around.
 Fig. \ref{fig2Dweff}
 plotted  from Eq. (\ref{Phantomweffkt}) setting $t_a=1$ and
 $t=0.7$ shows that if $k=\pm 1$, $q$ could be negative, i.e. phantom
 accelerated expansion, even when $w_{\rm eff} > -1$.
 Regardless of $t_a$ and $t$,  %
 \bea \lim_{q\rightarrow -\infty} w_{\rm eff}(q)
= -1\, \;\;\;\; {\rm and} \;\;\;\;  \lim_{q\rightarrow +\infty}
w_{\rm eff}(q) =
 -\f{1}{3}\,, \eea for phantom expansion.  In particular, for $k=-1$,
$w_{\rm eff}>0$ could give $q<0$ and $w_{\rm eff}$ is infinite when
$\ln{q}/\ln{(t_{\rm a}-t)} + q =1$ (see Eq. (\ref{locus})).

\section{Scalar field exact solution} \label{sec:solution}
\subsection{Bound value of $\phi(t)$ from effective equation of state for $k=0$ case}%
 In flat universe, the phantom expansion occurs when $w_{\rm eff} <
-1$. Using Eqs. (\ref{phanp}), (\ref{wphi}) in Eq. (\ref{weff}), we
get a bound \bea \epsilon \dot{\phi}^2 &<& -
\frac{n}{3}\rho_{\gamma} \,. \eea
Assuming $a(t) = {(t_{\rm a}-t)^{q}}$ and phantom scalar field, i.e.
$\epsilon=-1$ with using Eq. (\ref{barorho}), the solution is found
to be in the region,
\bea \phi(t) > \frac{1}{\beta}\sqrt{\frac{Dn}{3}}\,(t_{\rm a}
-t)^{-\beta} + \phi_0\,. \label{phantomphisol} \eea where $\beta
\equiv (qn-2)/2\,. $
\subsection{Solution solved from Friedmann equation}
\subsubsection{Scalar field potential in flat and scalar field
dominated case} A simplest case for analysis is when considering
flat universe ($k=0$) with negligible amount of barotropic fluid
($D=0$). The Eq. (\ref{phit})
 is hence simply integrated out. The solution is %%%%
\be %%%
\phi(t) = \pm \frac{1}{\kappa}\sqrt{\frac{2q}{\epsilon}}\, \ln (t_{\rm a} -t) \,+\,\phi_0 %
 \ee
Insert this result into Eq. (\ref{vtphantom}), we obtain the scalar
field potential,%%%%
\be %%%
V(\phi) = \frac{q(3q-1)}{\kappa^2} \exp \left\{  \pm \kappa \sqrt{\frac{2\epsilon}{q}} [\phi(t) -\phi_0] \right\}\,.  %
 \ee
The solutions above are real only when $q$ and $\epsilon$ have the
same sign, i.e. when $\epsilon = 1, q > 0$ and $\epsilon = -1, q <
0$. This looks similar to potential that gives power-law expansion
as well-known \cite{Lucchin}. It is not surprised since in our case
($q<0$) it has been known that phantom field, when rolling up the
hill of slope-varying exponential potential (varying $q$), results
in phantom expansion $a \sim (t_{\rm a}-t)^q$
\cite{Caldwell:2003vq}.

\subsubsection{Solution for $k=0$, $D \neq 0$ case}
For the case $k=0$ with $D \neq 0$, the solution of Eq. (\ref{phit})
is %%
\bea %
\phi(t) & = & \pm \f{1}{qn-2} \sqrt{\f{2q}{\epsilon \kappa^2}}
\times \nonumber \\ & &\left\{ \ln \left[
 \f{(t_{\rm a}-t)^{-qn+2}}{\left(1 + \sqrt{1-(nD\kappa^2/6q)(t_{\rm a}-t)^{-qn+2}}\right)^2}  \right]
 + 2\sqrt{1-\f{nD\kappa^2 (t_{\rm a}-t)^{-qn+2}}{6q}} +
\ln\l(\f{-nD\kappa^2}{6q}\r) \right\} +\phi_{0}\,, \label{phantomK0FriSol}\nonumber \\
\eea which is infinite when $q = 2/n$. The last logarithmic term in
the bracket is an integrating constant. Logarithmic function is
valid only when $q<0$.
\subsubsection{Solution for $k \neq 0$, $D = 0$ case}
\label{subsubsec:solfr}
For the reverse case, $k \neq 0, D = 0$, the solution is
\bea \phi(t) & = & \pm \f{1}{q-1} \sqrt{\f{2q}{\epsilon \kappa^2}}
\times
 \nonumber \\ &  &\left\{ \ln \left[
 \f{(t_{\rm a}-t)^{q-1}}{\sqrt{k/q}}\left(1+ \sqrt{\l(\f{k}{q}\r)(t_{\rm a}-t)^{-2q+2} + 1} \: \right)  \right]
 - \sqrt{\l(\f{k}{q}\r)(t_{\rm a}-t)^{-2q+2} + 1}\:  \right\} +
\phi_0\,, \no \\ %
\eea %
which becomes infinite when $q=1$. The values $q$ and $\epsilon$
must have the same sign for it to be real-value function. The case
$k\neq 0$ with $D\neq 0$ can not be found analytically except when
setting $n=2\;\, (w_{\gamma}=-1/3)$ which is not natural fluid.

\subsection{Solution solved with NLS-type formulation}
One can obtain exact solution of Eq. (\ref{phit}) indirectly via
NLS-type formulation. Consider Eq. (\ref{Pxphantom}), we notice that
setting $D=0$ does not make sense in NLS-formulation since even $D$
vanishes, $n$ (barotropic fluid parameter) still appears in other
terms. Therefore we can only consider non-zero $D$ case. Assuming $k
= 0$ with $D\neq 0$ and using Eq. (\ref{Pxphantom}) in Eq. (\ref{psi}), the solution is %
\bea %
\psi(x) &=& \pm  \s{\f{8q}{\e \k^2 (qn-2)^2}} \times \nonumber
\\ & & \l\{ - \s{1 - \l[  \f{\k^2 D n(qn-2)^2}{24q}
(x-x_0)^2 \r]} + \ln \l[ \f{1+ {\s{1 - \l[ {\k^2 D n (qn-2)^2}/{24
q}\r] (x-x_0)^2 }}}{ (x-x_0)} \f{4qn}{\e (qn-2)^2}  \r] \r\}\,. \no
\\ %
\eea
Transforming to $t$ variable using Eq. (\ref{xast}),
\bea %
\phi(t) & = & \pm \f{1}{qn-2} \sqrt{\f{2q}{\epsilon \kappa^2}}
\times \no \\ & & \left\{ \ln \left[
 \f{(t_{\rm a}-t)^{-qn+2}}{\left(1 + \sqrt{1-(nD\kappa^2/6q)(t_{\rm a}-t)^{-qn+2}}\right)^2}  \right]
+ 2\sqrt{1-\f{nD\kappa^2 (t_{\rm a}-t)^{-qn+2}}{6q}}  +
\ln\l(\f{qn-2}{2qn}\r)^2 \right\} +\phi_{0}\,. \label{PhantomK0NLSSol} \nonumber \\
\eea
The only difference from the solution (\ref{phantomK0FriSol})
obtained from standard method is the logarithmic integrating
constant term in the bracket. In case of $k \neq 0$ with $D\neq 0$,
the integral (\ref{psi}) can not be integrated analytically even
when assuming $n$ value except for $n = 2$ which is integrable.
However $n=2$ is not natural fluid. This is similar to using
standard method in Sec. \ref{subsubsec:solfr}.

\section{Conclusions} \label{sec:conclusion}
We consider a system of FLRW cosmology of scalar field and
barotropic fluid assuming phantom acceleration. We have worked out
cosmological quantities in the NLS-formulation of the system for
flat and non-flat curvature. The Schr\"{o}dinger wave functions are
illustrated in Fig. \ref{figux} for various types of barotropic
fluid. These wave functions are non-normalizable. We show
Schr\"{o}dinger potential plots for dust and radiation cases in
closed, flat and open universe. The procedure considered here is
reverse to a problem solving in quantum mechanics in which the
Schr\"{o}dinger potential must be known before solving for wave
function. In NLS formulation, the Schr\"{o}dinger equation is
non-linear (reducible to linear in some cases) and the wave function
is expressed first by the expansion function, $a(t)$. Afterward the
Schr\"{o}dinger potential is worked out based on expansion function
assumed. Moreover, the NLS total energy $E$ is negative (see Eq.
(\ref{E})). We also perform analysis on effective equation of state.
We expresses $w_{\rm eff}$ in term of $q$ and $k$. In a non-flat
universe, there is no fixed $w_{\rm eff}$ value for a phantom
divide. We show this by analyzing Eq. (\ref{Phantomweffkt}) and by
presenting illustrations in Figs. (\ref{figweff}) and
(\ref{fig2Dweff}). In these plots, even $w_{\rm eff} > -1$, the
expansion can still be phantom, i.e. $q$ can be negative.
Especially, in $k=-1$ case, positive $w_{\rm eff}$ could also give
$q<0$. The value of $w_{\rm eff}$ approaches -1 when $q \rightarrow
-\infty$ and $-1/3$ when $q\rightarrow +\infty$. In open universe,
$w_{\rm eff}$ blows up when $\ln{q}/\ln{(t_{\rm a}-t)} + q =1$.

The last part of this work is to solve for scalar field exact
solution for phantom expansion. Within framework of the standard
Friedmann formulation, we obtained exact solution in simplest case
where scalar field is dominated in flat universe. Apart from that we
also obtained exact solutions in the cases of non-flat universe with
scalar field domination and flat universe with mixture of barotropic
fluid and scalar field. Afterward, we use NLS formulation, in which
the wave function is equivalent to the scalar field exact solution,
to solve for the exact solutions.  We can apply the NLS method to
solve for the solution only when the barotropic fluid density is
non-negligible. Setting $D=0$ in NLS framework is not sensible
because even if $D$ term vanishes, the barotropic fluid parameter
$n$ still appears in other terms of the wave function. This is a
disadvantage point of the NLS formulation.

Transforming standard Friedmann formulation to NLS formulation
renders a few effects to the integration. In standard form (Eq.
(\ref{phit})), $n$ appears in only $D$-term and all terms are
$t$-dependent. In NLS-form (Eq. (\ref{Pxphantom}) when inserted in
Eq. (\ref{psi})), $D$-term becomes a constant ($E$), hence the
number of $x$(or equivalently $t$)-dependent terms is reduced by
one. This is a good aspect of the NLS. In both Friedmann-form and
NLS-form, the solutions when $k\neq 0 $ and $D\neq 0$ are difficult
or might be impossible to integrate unless assuming values of $q$
and $n$. Therefore reduction number of $x$-dependent term helps
simplifying the integration.

\section*{Acknowledgements}
T.~P. is sponsored by the Thailand Riean Dee Science Studentship. R. S. is supported via the Tah Poe Academia Institute's Research Assistantship
under the Naresuan Faculty of Science Research Scheme. B.~G. is a TRF Research Scholar under the Thailand Research Fund. B. G. expresses his
gratitude to Naresuan University's Overseas Visiting Postdoctoral Research Fellowship and to Anne-Christine Davis, Stephen Hawking and Neil
Turok for providing additional support at the Centre for Theoretical Cosmology, D.A.M.T.P., University of Cambridge.

\end{document}